\documentclass[conference]{IEEEtran}
\IEEEoverridecommandlockouts
\usepackage{cite}
\usepackage{amsmath,amssymb,amsfonts}
\usepackage{algorithmic}
\usepackage{graphicx}
\usepackage{textcomp}
\usepackage{xcolor}
\usepackage{color}
\usepackage{tcolorbox}
\usepackage{multirow}
\def\BibTeX{{\rm B\kern-.05em{\sc i\kern-.025em b}\kern-.08em
    T\kern-.1667em\lower.7ex\hbox{E}\kern-.125emX}}

\usepackage{changes}
\definechangesauthor[name={Tanghaoran Zhang}, color=blue]{zthr}
\definechangesauthor[name={Zhang Zhang}, color=pink]{ZDouble}

\newcommand{\cmmnt}[1]{}

\newcommand*{\ie}{i.e., }

\newboolean{showcomments}
\setboolean{showcomments}{true}
\ifthenelse{\boolean{showcomments}}
 { \newcommand{\mynote}[2]{
      \fbox{\bfseries\sffamily\scriptsize#1}
        {\small$\blacktriangleright$\textsf{\emph{#2}}$\blacktriangleleft$}}}
        { \newcommand{\mynote}[2]{}}

\usepackage{hyperref}
\begin{document}
\title{An Empirical Study on Distilling ChatGPT for Advancing Code
Intelligence Tasks\\

}

\author{

\IEEEauthorblockN{
        Kang Yang\IEEEauthorrefmark{1},
        Xinjun Mao\IEEEauthorrefmark{1},
        Shangwen Wang\IEEEauthorrefmark{1},
        Tanghaoran Zhang\IEEEauthorrefmark{1},
        Bo Lin\IEEEauthorrefmark{1},
        Yanlin Wang\IEEEauthorrefmark{2},\\
        Yihao Qin\IEEEauthorrefmark{1},
        Zhang Zhang\IEEEauthorrefmark{1},
        Xiaoguang Mao\IEEEauthorrefmark{1}
    }

\IEEEauthorblockA{\IEEEauthorrefmark{1} National University of Defense Technology \\
    \{yangkang, xjmao, wangshangwen13, zhangthr, linbo19, qinyihao, zhangzhang14, xgmao\}@nudt.edu.cn
    \IEEEauthorblockA{\IEEEauthorrefmark{2} Sun Yat-sen University, wangylin36@mail.sysu.edu.cn}
    }
}


\maketitle
\begin{abstract}
Pre-trained code models have emerged as crucial tools in various code intelligence tasks. However, their effectiveness heavily depends on the quality of the pre-training dataset, particularly the human reference comments, which usually serve as a bridge between the programming language and natural language. One significant challenge is that such comments can become inconsistent with the corresponding code as the software evolves.
This discrepancy can lead to suboptimal training of the models, decreasing their performances. 

Large language models (LLMs) have recently demonstrated superior capabilities in generating high-quality code comments. In light of that, we try to tackle the quality issue of the dataset by harnessing the power of LLMs. Specifically, we raise the question: {\em Can we rebuild the pre-training dataset by substituting the original comments with LLM-generated ones for more effective pre-trained code models?}

To answer the question, we first conduct a comprehensive evaluation to compare ChatGPT-generated comments with human reference comments. 
As existing reference-based metrics treat the reference comments as gold standards, we introduce two auxiliary tasks as novel reference-free metrics to assess the quality of comments, \ie code-comment inconsistency detection and code search.
Experimental results show that ChatGPT-generated comments demonstrate superior semantic consistency with the code compared to human reference comments, indicating the potential of utilizing ChatGPT to enhance the quality of the pre-training dataset. 

Based on this finding, we rebuilt the widely used dataset, CodeSearchNet, with ChatGPT-generated comments. Subsequent experiments involve re-pre-training the widely used model CodeT5 with our refined dataset.Evaluation results on four generation tasks and one understanding code intelligence tasks
show that the model pre-trained by ChatGPT-enhanced data outperforms its counterpart (pre-trained by original human reference comments data) on code summarization, code generation, and code translation tasks.
This research validates the feasibility of rebuilding the pre-training dataset by ChatGPT for advancing code intelligence tasks and advocates rethinking the reliance on human reference comments for code-related tasks.


\end{abstract}

\section{Introduction}
\label{sec:introduction}
In the realm of code intelligence, pre-trained code models have significantly enhanced a spectrum of tasks such as code summarization~\cite{tenny1988program, woodfield1981effect, song2019survey}, code generation~\cite{xu2020incorporating, singh2023codefusion, shin2023good}, code search~\cite{gu2018deep, sun2022importance, xie2023survey}, clone detection~\cite{svajlenko2014towards, lei2022deep} and automated bug fixing~\cite{li2020dlfix, goues2019automated, mamatha2022literature}. Pre-trained source code models such as CodeBERT~\cite{feng2020codebert}, GraphCodeBERT~\cite{guo2020graphcodebert},
CodeT5~\cite{wang2021codet5}, and UniXcoder~\cite{guo2022unixcoder} and have achieved state-of-the-art (SOTA) results on various software engineering tasks. 
The comments of the corresponding code snippets serve as a crucial bridge between the programming language (PL) and natural language (NL), providing contextual understanding pivots that are vital for pre-training the above models.
As such, the effectiveness of the pre-trained code models relies heavily on the quality of their pre-training datasets~\cite{sun2022importance}, which traditionally depend heavily on human reference comments.

However, this reliance on comments introduces a fundamental challenge: as software evolves, these comments often become outdated or mismatched with the code~\cite{steidl2013quality, ying2005source, wen2019large}, leading to semantic inconsistencies between the code and comment.
Previous research has highlighted such inconsistencies. 
Shi et al.~\cite{shi2022we} revealed that 41.9\% of the code summarization dataset TLC~\cite{hu27summarizing} is noisy, with 22.8\% of the comments not matching the corresponding code snippets. Consequently, inconsistent comments deteriorate the quality of the PL-NL dataset, which can degrade the training efficacy and performance of code models. Sun et al.~\cite{sun2022importance} identified that more than one-third of the comments in CodeSearchNet (Java)~\cite{husain2019codesearchnet} did not describe core functionalities, and the model trained with noisy data faces severe performance degradation~\cite{sun2022importance}.

Recently, large language models (LLMs) like ChatGPT have demonstrated superior generation capabilities~\cite{guo2023exploring, liang2023large}, especially in generating high-quality code comments~\cite{geng2024large}. Specifically, the results show that when given adequate examples, the effectiveness of LLMs would be significantly higher than traditional deep learning-based approaches. 
In light of this, we are motivated to utilize LLMs to address the limitations faced by pre-trained code models. This paper aims to answer the following question: \textit{\textbf{Can we rebuild the pre-training dataset by substituting the original comments with LLM-generated ones for more effective pre-trained code models?}}

To that end, we first conduct a comprehensive evaluation to compare LLM-generated comments with human reference comments. 
We investigate two types of LLMs published by OpenAI in this evaluation, including gpt-3.5-turbo and text-davinci-003. 
Traditional evaluation metrics for code summarization are reference-based, focusing on measuring the similarity between predicted comments and reference comments. 
However, the underlying assumption of the reference-based evaluation is that reference comments are gold standard superior to other baselines, whether n-gram overlap or semantic similarity measurements are utilized. 
A critical purpose of this work is to compare the quality of ChatGPT-generated comments with human-written reference comments; thus, a reference-free evaluation is required to enable the comparison.

In our study, we introduce two auxiliary tasks – code-comment inconsistency detection and semantic code search – to offer a more refined reference-free assessment of the semantic similarity between the code and its associated comments. 
The inconsistency detection task aims to identify inconsistency between comment and code, and the code search task assesses the ability to retrieve the correct code snippet using its comment as a query. 
Our intuition is that a higher-quality comment would exhibit better semantic consistency with the corresponding code so that (1) it is less likely to be detected as inconsistent by a well-trained classifier and (2) it is expected to facilitate accurate retrieval of the associated code from a database when used as a search query. 
Our experimental results show that \textit{\textbf{comments generated by ChatGPT preserve better semantic consistency with the code than human reference comments}}. 

This insight forms the basis to reconstruct the popularly utilized CodeSearchNet~\cite{husain2019codesearchnet} dataset, substituting the human reference comments with ChatPT-generated ones. We further extend our research by pre-training the widely used model CodeT5~\cite{wang2021codet5} with this updated dataset and evaluating its performance across five downstream code intelligence tasks. 
The results of these experiments are revealing. 
The model trained with the ChatGPT-enhanced dataset exhibit superior performance in code intelligence tasks like code summarization, code generation, and code translation compared to the counterpart model trained with original datasets containing human reference comments. These findings affirm the potential of LLMs in improving the quality of training data for code intelligence tasks and underscore the need to reevaluate the longstanding reliance on human reference comments. 

In summary, this research makes the following contributions: 
\begin{itemize}
    \item Innovative Evaluation Metrics: To the best of our knowledge, we are the first to adopt code comment inconsistency detection and code search to automatically evaluate the quality of code comments, which offer a reference-free approach to evaluating the quality of comments generated by ChatGPT as well as written by human. 
    \item Empirical Validation of ChatGPT: Our experimental results demonstrate that ChatGPT-generated comments achieve better semantic consistency with code than human reference comments. This finding validates the quality and reliability of ChatGPT-generated comments in the context of software engineering.
    \item Exploration of ChatGPT in Dataset Construction: 
    Our work explores using ChatGPT to create high-quality code comment datasets, demonstrating its practicality and effectiveness in training source code models and advancing code intelligence tasks. 
    In addition, we rebuilt the large-scale CodeSearchNet dataset with +2M high-quality comments generated by ChatGPT.
\end{itemize}

\section{Preliminaries}
\label{sec:preliminaries}
This section reviews the NL-PL pre-trained models, as well as the evaluation of code summarization.
\subsection{NL-PL Pre-trained Models}
A standard NL-PL pre-trained model involves initially training a large-scale model using self-supervised objectives on extensive unlabelled datasets, followed by fine-tuning it for specific downstream applications, such as tasks related to code understanding and generation, using task-specific loss functions. This paradigm, first introduced in natural language processing (NLP) \cite{kenton2019bert, radford2018improving}, has been adapted for programming tasks. Many NL-PL pre-trained models \cite{feng2020codebert, guo2020graphcodebert, wang2021codet5} have been proposed with their typical architectures and pre-training tasks. 

CuBERT \cite{kanade2020learning} and CodeBERT \cite{feng2020codebert} are the pioneer models, and the CodeBERT is the first large NL-PL pre-trained model for multiple programming languages. Following BERT \cite{kenton2019bert} and RoBERTa \cite{liu2019roberta}, multi-layer bidirectional Transformer \cite{vaswani2017attention} are utilized as the model architecture. CodeBERT is pre-trained on a bimodal dataset CodeSearchNet \cite{husain2019codesearchnet}, and two objectives, Masked Language Modeling (MLM) \cite{kenton2019bert} and Replaced Token Detection (RTD) \cite{clark2020electra}, are employed as pre-training tasks. GraphCodeBERT \cite{guo2020graphcodebert}, sharing CodeBERT's architecture, goes a step further by incorporating the structural aspects of code, specifically the data flow graph (DFG). While maintaining the MLM objective, GraphCodeBERT does away with the RTD objective. Instead, it introduces two DFG-related tasks: predicting data flow edges and aligning nodes. In four downstream tasks, GraphCodeBERT has been shown to surpass CodeBERT's performance \cite{guo2020graphcodebert}. 

The encoder-only based models necessitate an additional decoder for generation tasks, which cannot leverage the advantages of the pre-training phase. On the contrary, GPT-C \cite{svyatkovskiy2020intellicode} and CodeGPT \cite{lu2021codexglue} are pre-trained by using unidirectional language modeling, which is good at auto-regressive tasks like code generation, but is sub-optimal for code understanding tasks.

Recent studies have focused on encoder-decoder models to accommodate understanding and generation tasks. For instance, PLBART \cite{ahmad2021unified} utilizes the BART architecture \cite{lewis2020bart} and is pre-trained on both NL and PL with denoising objectives. CodeT5 \cite{wang2021codet5}, developed by Wang et al., modifies the T5 model \cite{raffel2020exploring} to incorporate essential token type information from identifiers, facilitating multi-task learning in downstream applications. 

In contrast to the prevailing focus in current research on optimizing model architectures and designing domain-specific pre-training tasks, our study shifts the spotlight to another critical factor influencing model performance: the quality of the pre-training data. This work investigates how enhancements in the pre-training dataset quality can influence the performance of pre-trained code models on downstream code intelligence tasks.

\subsection{Evaluation of Code Summarization}

Quantitatively evaluating the quality of the generated summary is a non-trivial task. Typically, this evaluation involves comparing the model-generated summary with a reference summary, which is often the original comment provided by a developer.
These reference summaries serve as a crucial benchmark against which the generated summaries are assessed. The degree of similarity between the predicted summary and the reference summary is then calculated to evaluate the extent to which the predicted summary aligns with the reference, thus serving as an indicator of the quality of the predicted summary. These metrics are called reference-based ones and are typically classified into two main categories: n-gram overlap and semantic similarity \cite{zhang2019bertscore, haque2022semantic} in research communities.

The dominant evaluation methods are traditional n-gram overlap-based metrics like BLEU \cite{papineni2002bleu}, which were originally utilized in the machine translation community for measuring the predicted translations’ similarity to reference translations.  BLEU evaluates the similarity between the predicted and reference summaries by comparing n-grams within them. Commonly, $n$ varies from 1 to 4, allowing for the calculation of a $BLEU_n$ score to gauge performance:
\begin{equation}
\label{formula:bleu}
  BLEU_n = \frac{\sum_{t_n} \min \{ Cp(t_n), Cr(t_n) \}}{P(n)}
\end{equation}
In this formula, $BLEU_n$ represents the BLEU score for n-grams, $Cp(t_n)$ is the count of n-grams in the predicted summary, $Cr(t_n)$ is the count of n-grams in the reference summary, and $P(n)$ is the total number of n-grams in the predicted summary.

A limitation of BLEU is its design as a corpus-level measure rather than a sentence-level one. A comprehensive review revealed that BLEU's correlation with human judgment is notable only at the corpus level \cite{reiter2018structured}. However, programmers typically meet individual code summaries in practice. This disparity in BLEU's applicability at the sentence level has been a point of contention among scholars across various disciplines \cite{van2019best, novikova2017we}. Although ROUGE \cite{lin2004rouge} and METEOR \cite{banerjee2005meteor}) have been proposed to address specific complaints about BLEU, they are word/n-gram overlap-based metrics. Essentially, they still compute literally whether the same words/n-gram appear in the same order in both predictions and references.

Measuring semantic similarity, as an alternative to word overlap methods, assesses word similarity within an embedding space, providing ``partial credit'' for word matches, as termed by Weiting et al. \cite{wieting2019beyond}. Rather than binary scoring, semantic similarity based metrics apply a graded score that reflects the semantic proximity of words in a pre-trained embedding. The application of semantic similarity to code summary assessment is supported by growing evidence of the insufficiency of mere word overlap \cite{stapleton2020human, mahmud2021code}. 
Mahmud et al. advocate for BERTScore as it may more effectively capture semantic similarities than traditional symbolic similarities \cite{mahmud2021code}. Haque et al. explore word-based and sentence-based semantic similarity metrics in assessing code comment summarization. 
They conclude that the cosine similarity derived from the Sentence-BERT Encoder \cite{reimers2019sentence} and Universal Sentence Encoder \cite{cer2018universal} sentence representations most closely aligns with the similarity perceived by human evaluators. 

Reference-based metrics, including n-gram overlap and semantic similarity based ones, are traditionally used on the assumption that human-written comments serve as a benchmark for quality. However, these metrics are primarily designed to assess the similarity between machine-generated comments and these human-written references rather than directly comparing the quality of one against the other. 
In the context of LLMs, which can generate comments with high levels of proficiency, these metrics may not fully capture the nuances of quality, especially when it comes to determining whether the machine-generated content is of equal or superior quality compared to the human references.

\section{Evaluation of ChatGPT-generated comments}
\label{sec:evaluation}
\subsection{Objective and Research Questions}

This study investigates the potential of ChatGPT to rebuild training datasets for advancing code intelligence tasks. Before constructing the dataset, we focus on the comprehensive comparison of ChatGPT-generated comments and human reference comments in this section. 
Traditional evaluation approaches, such as n-gram overlap based metrics and semantic similarity based metrics, are reference-based evaluation approaches. However, these reference-based metrics cannot directly compare the quality of ChatGPT-generated comments and human reference comments because references are seen as gold standard. Recognizing this gap, our study proposes a paradigm shift in evaluation strategies. We focus on the practical utility and real-world applicability of the comments, moving beyond traditional metrics to extrinsic evaluation tasks.
To address this, we utilize two code intelligence tasks: code-comment inconsistency detection and semantic code search, which effectively measure the semantic alignment between the code and its corresponding comments. Thus, 
these two tasks are utilized as reference-free metrics for extrinsic comment evaluation. 
Guided by the two principal research questions (RQs), our study explores:

\textbf{RQ1}: \textbf{How effective are the proposed reference-free metrics in assessing the quality of code comments? } 
This question explores the effectiveness of our newly proposed evaluation metrics and examines their alignment with traditional reference-based metrics, which lays the foundation for a comparative evaluation of ChatGPT-generated comments and human reference comments.

\textbf{RQ2: How does the quality of comments generated by ChatGPT compare to human-written comments?}
This question investigates the direct quality comparison of ChatGPT-generated comments and human reference comments, which is central to determining the practicality of employing ChatGPT to enhance training datasets for code intelligence tasks.

\subsection{Reference-free Evaluation Metrics}
\subsubsection{Code Comment Inconsistency Detection}
\label{ssbsec:ccid}
The task of code comment inconsistency detection (CCID) is to determine whether a comment is semantically misaligned with the corresponding code snippet \cite{wen2019large}. Since CCID is of immense practical use to software developers who have a vested interest in keeping their code bases easily readable, navigable, and as bug-free as possible, prior works proposed kinds of approaches for detecting inconsistencies \cite{ panthaplackel2021deep, steiner2022code}. 

\begin{figure}
\centerline{\includegraphics[scale=0.50]{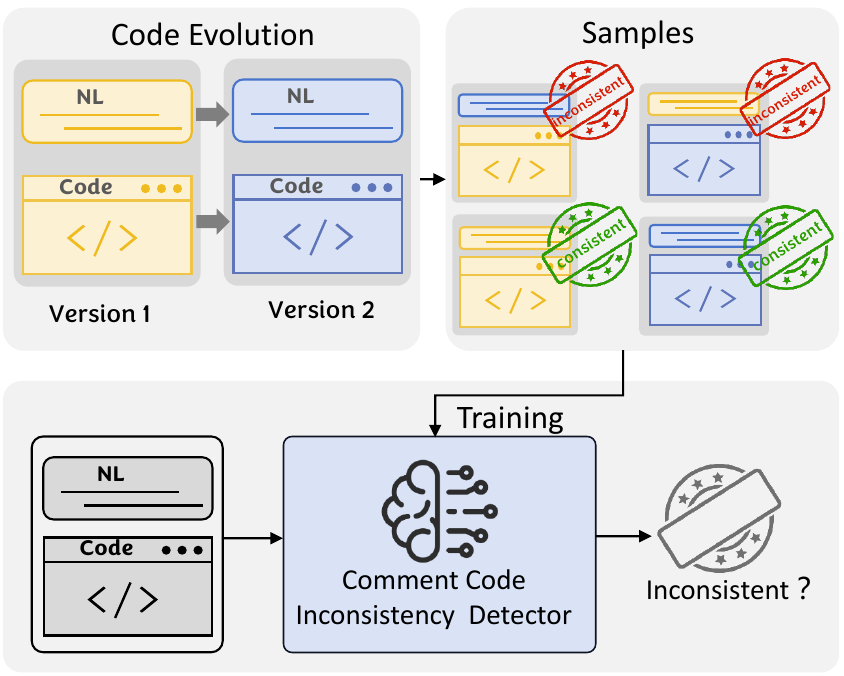}}
\caption{ Code comment inconsistency detection as the reference-free evaluation metric.}
\label{fig-inconsistency}
\end{figure}

Our motivation stems from the hypothesis that the most informative comments are those that accurately mirror their corresponding code. Inconsistent comments can cause confusion, errors, or misinterpretations, reducing effectiveness. The CCID task is designed to assess the consistency between comments and their code quantitatively. This enables us to identify discrepancies that may indicate outdated or erroneous comments, providing a precise and objective means to evaluate comment quality. Furthermore, the ability to detect inconsistency is intrinsically linked to a deeper understanding of the code-comment relationship and can serve as a valuable tool for software maintenance. By focusing on the alignment of code and comments, we can offer insights into the utility of comments and their role in enhancing the clarity and correctness of software documentation. Therefore, we introduce the CCID task to evaluate the quality of code comments by measuring the inconsistency rate in test datasets.

The data we used to train the CCID classifier is curated by Panthaplackel et al.~\cite{panthaplackel2021deep}~\footnote{https://drive.google.com/drive/folders/1heqEQGZHgO6gZzCjuQD1EyYert\\N4SAYZ}, which includes 40,688 samples of @return, @param, and summary Javadoc comments paired with their corresponding Java code methods. They consider comment-code pairs from each version of consecutive commits: $(c_1, nl_1), (c_2, nl_2)$. 
We collect examples that code changes do exist between two versions in which $c_1$ $\neq$ $c_2$. 
As a result of code changes, the developer updated the comment because it would have otherwise become inconsistent. 
Therefore, if $nl_1$ $\neq$ $nl_2$, we take $nl_1$ comment to be inconsistent with $c_2$ code. As illustrated in Figure ~\ref{fig-inconsistency}, $(c_1, nl_2)$ and $(c_2, nl_1)$ are consequently constitute the inconsistent positive examples. 
In contrast, if $nl_1$ = $nl_2$, the collected examples (i.e. $(c_1, nl_1)$ and $(c_2, nl_2)$) are labelled as consistent negative examples. The assumption behind this is that the developer chose not to update the comment while modifying the code, as the comment was still consistent with the changes~\cite{panthaplackel2021deep}. 
Figure~\ref{fig-inconsistency} demonstrates the data constructing and training process of the CCID classifier. 

Nevertheless, this procedure of collecting data brings in inaccuracies~\cite{panthaplackel2021deep}. If a comment is refined without changing its semantics, there are positive mislabelled examples. Hence, even though $nl_1$ $\neq$ $nl_2$ and $c_1$ = $c_2$, the comment is actually consistent with code.
To alleviate this noise, we empirically consider the $nl_1$ $\neq$ $nl_2$ examples that there is a difference of more than 30\% in tokens between them in our work. The assumption is that 30\% tokens' difference largely results in semantic changes in comments evolution.

Suppose $f$ denotes the CCID classifier, and the input of $f$ is a code snippet $c$ and its corresponding comment $nl$. The output $f(c, nl) = 1$ indicates they are semantically inconsistent; otherwise, $f(c, nl) = 0$. Assume $<C, NL>$ is the code snippets, and its paired model-generated/human-written comments in test datasets, and the total examples in test datasets is $N$. We define the inconsistency rate (\textit{IncRate}) by calculating the proportion of inconsistent examples. 
\begin{equation}
  IncRate=\frac{1}{N} \sum_{i=0}^{N}f(c_i,nl_i)
\end{equation}

We propose to use \textit{IncRate} as a reference-free metric to evaluate the quality of comments, and the lower \textit{IncRate} indicates better semantic consistency between the code and the comments.  

\subsubsection{Semantic Code Search}
\label{ssbsec:codesearch}
The semantic code search task is to retrieve a code snippet that matches a given query by effectively capturing the semantic similarity between the query and code. Semantic code search is a vital software development assistant that significantly improves development efficiency and quality. 

Our motivation is rooted in the premise that the code search task could serve as an indicator of the degree to which comments are aligned with their corresponding code.
We assume that high-quality comments should not only elucidate code functionality but also enhance the discoverability of code snippets through semantic code search. 
This assumption aligns with the practical use case of developers who regularly rely on comments to navigate and understand large code bases. Additionally, code comments are used as an alternative to the practical queries in research communities~\cite{husain2019codesearchnet, sun2022importance}. Consequently, the effectiveness of comments in facilitating accurate and efficient code search results serves as a proxy for their quality, providing a concrete, measurable dimension to an otherwise subjective attribute of software documentation. Therefore, we argue comments that are more helpful for code search tasks are likely to be of higher quality. In our study, we employ comments as queries within a code search task to assess the comment quality.

\begin{figure}
\centerline{\includegraphics[scale=0.55]{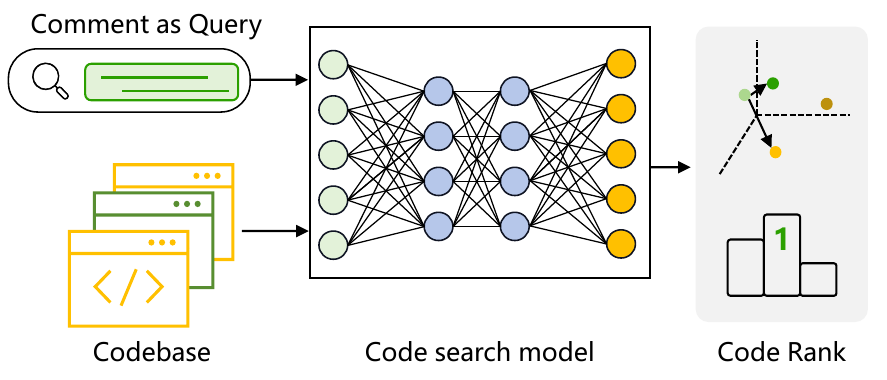}}
\caption{ Code search as reference-free evaluation metric.}
\label{fig-codesearch}
\end{figure}

To evaluate the performance of code search, we use the Mean of Reciprocal Rank (MRR) \cite{gu2018deep,lv2015codehow,ye2014learning}, which has been widely adopted in the evaluation of semantic code search. 
The $MRR$ score quantifies the ranking of the target code snippet to the given comment query, and it only cares about where the most relevant result is ranked. 
We use CodeBERT~\cite{feng2020codebert} for code search task, and follow the configuration reported in their artifacts~\footnote{https://github.com/microsoft/CodeBERT/tree/master/CodeBERT/codesearch} to construct data and fine-tune model. When computing $MRR$ scores in the testing set, for each query-code pair, $|Q|-1$ code snippets from other pairs in the same batch play the role of distractor codes, where $|Q|$ is the batch size, and we set as $1000$ in this work. For a single batch, $MRR$ is calculated as follows: 

\begin{equation}
  MRR = \frac{1}{\left | Q \right | } \sum_{1}^{\left | Q \right | } \frac{1}{Rank(\tilde{c}_i,nl_{i} )} 
\end{equation}
\noindent where $\tilde{c}_i$ is the ground-truth code snippet for its paired comment query $nl_{i}$, and $Rank(\tilde{c}_i,nl_{i} )$ is its corresponding rank in the retrieved results. $MRR$ gives a score of the predicted result based on its rank. The average value of all batches is the final $MRR$ score. 

We propose to use $MRR$ as a reference-free metric to evaluate the quality of comments. A higher $MRR$ indicates better semantic consistency between the code and the comments. 

\subsection{Experiment Design}
\subsubsection{Dataset}
We conduct experiments on a widely used Java benchmark dataset TLC \cite{hu27summarizing}, which has 66k code-comment pairs collected from more than 9K open-source Java projects created from 2015 to 2016 with at least 20 stars. They first extracted Java methods and Javadocs and treated the first sentence of the Javadoc as the ground-truth comment of the corresponding code. We directly use the TLC dataset open-sourced by the previous work \cite{shi2022we}. The training/validation/test set contains 53,528/7,555/4,985 samples, respectively.
\subsubsection{Baselines and ChatGPT}
In our research, we aim to conduct a comprehensive analysis of comment quality, comparing the performance of ChatGPT with that of human-written comments. To provide a meaningful context for evaluation, we introduce several deep learning (DL) based code summarization models as baselines. Specifically, we incorporate NCS \cite{ahmad2020transformer}, SIT \cite{wu2021code}, and DOME \cite{mu2023developer} as benchmarks to establish a foundation for assessing the quality of ChatGPT-generated comments and human reference comments. This comprehensive comparison will facilitate a more nuanced understanding of the performance differences in DL model-generated, ChatGPT-generated, and human-written comments, thereby enhancing the overall insight of our analysis.

\textbf{NCS} \cite{ahmad2020transformer} is a transformer-based approach that uses relative distances instead of absolute positions in the self-attention computation and applies a copy mechanism to copy rare tokens from the input source code. 

\textbf{SIT} \cite{wu2021code} is a structure-induced Transformer that integrates ASTs structure features into the self-attention calculation, which combines the AST tree, data-flow graph, and control-flow graph as a multi-view graph matrix to filter attention connections between tokens. 

\textbf{DOME} \cite{mu2023developer} is an approach that takes the comment intent information into account, which utilizes the intent-guided selective attention to explicitly select intent-relevant information from the source code, and produces various comments reflecting different intents. We use the same data split and follow the default training parameters reported in these papers.

\textbf{ChatGPT}~\footnote{https://openai.com/chatgpt} has become one of the most attention-grabbing achievements in the era of AI-generated content (AIGC). ChatGPT also has attracted widespread attention from software engineering as it exhibits excellent capability in software development. In this work, we investigate their performance in comment generation task, and two kinds of models published by OpenAI are taken into account. These two models are powerful enough and easy to access, leaving the time long enough for reproducing before it is deprecated. 
\begin{itemize}
    \item \textbf{gpt-3.5-turbo}. The most capable and cost-effective model in the GPT-3.5 family, which can understand and generate natural language and code. The snapshot of gpt-3.5-turbo from June 13th, 2023, gpt-3.5-turbo-0613 \footnote{https://platform.openai.com/docs/models/gpt-3-5}, is the version used in this work.
    \item \textbf{text-davinci-003}. The text-davinci-003 is the reinforcement learning with reward model trained from comparisons by humans, which is 10 times more expensive than gpt-3.5-turbo.
\end{itemize}

\begin{figure}
\centerline{\includegraphics[scale=0.40]{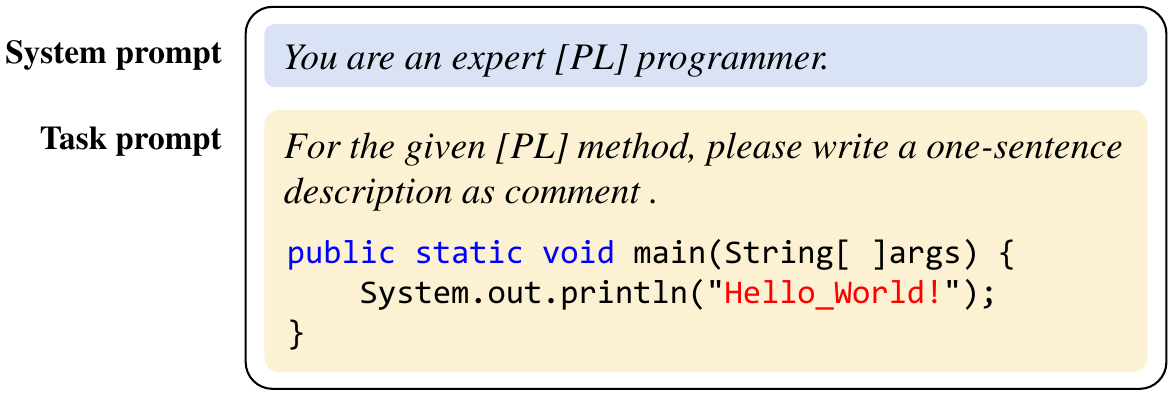}}
\caption{ The prompt design for code comment generation.}
\label{fig-prompt}
\end{figure}

To obtain summaries written by gpt-3.5-turbo/text-davinci-003, we design a simple prompt with the following format shown in Figure~\ref{fig-prompt}. 

\begin{table}
\centering
\caption{Statistics of the average length of words and the unique words appeared in human reference comments and ChatPT-generated comments on TLC dataset}
\label{tab:tlc-stat}
\begin{tabular}{ccc}
\hline
TLC dataset      & Avg.words length & Unique words \\ \hline
human references & 11.78            & 32,909        \\
gpt-3.5-turbo    & 13.38            & 40,619        \\
text-davinci-003 & 11.94            & 36,829        \\ \hline
\end{tabular}
\end{table}

We collected summaries from gpt-3.5-turbo/text-dacinci-003 using this prompt for a total of 66k Java methods, including 53,528 from the training set, 7,555 from the validation set, and 4,985 from the test set, respectively. As ChatGPT tends to give much longer comments than human-written references, we set the parameter \textit{max\_tokens=20} to meet the average length of human references when requesting OpenAI API for a fair comparison. Table~\ref{tab:tlc-stat} displays the statistics of word length and unique words for ChatGPT generated and reference comments. As for the sampling temperature parameters, we set the $top\_p=1$ and the $tempreture=1$.

\subsubsection{Evaluation Metrics}
\begin{figure}
\centerline{\includegraphics[scale=0.65]{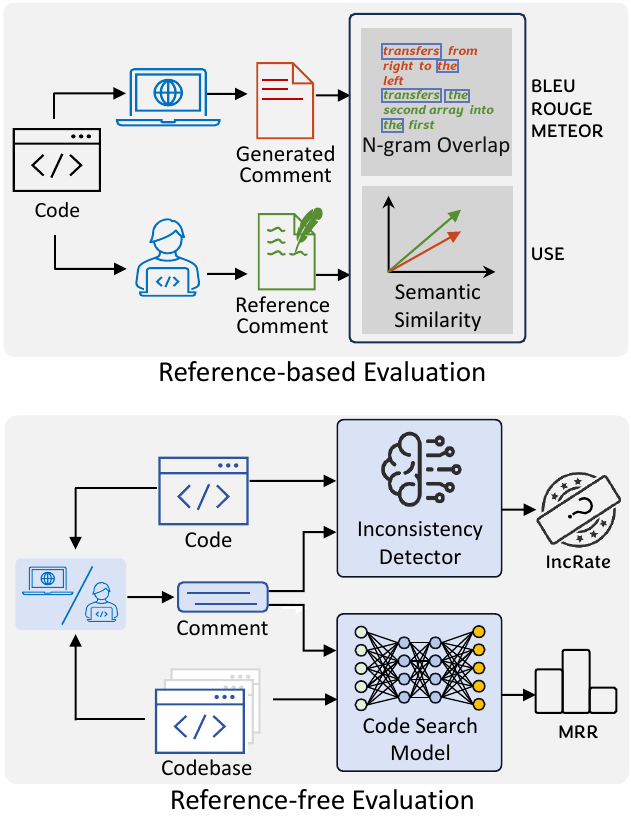}}
\caption{Reference-based and reference-free evaluation.}
\label{fig-ref-based-free}
\end{figure}

It is also necessary to report some commonly used traditional metrics in research communities for comparison as we propose new evaluation metrics. On one hand, this allows us to compare the consistency, strengths, and limitations of different metrics. On the other hand, it could offer effective evidence of our proposed methods when comparing the ChatGPT-generated comments with human reference comments. Therefore, we categorize the evaluation metrics into two main groups based on whether references are needed as a standard: (1) reference-free metrics and (2) reference-based metrics, as shown in Figure~\ref{fig-ref-based-free}.

\textbf{IncRate} is a reference-free metric for code summarization. As introduced in the former subsection~\ref{ssbsec:ccid}, IncRate evaluates the quality by measuring the percentage of inconsistent comment-code pair examples in the test set. 

\textbf{MRR} is originally used for the evaluation of information retrieval. As detailed in former subsection~\ref{ssbsec:codesearch}, we adopt it to measure the semantic relationship between the comment query and its paired code snippet. Higher MRR means a better quality of comment in semantic coherence.  

\textbf{USE} \cite{haque2022semantic} is a metric that encodes the reference and the predicted summary to a fixed-length vector using a universal encoder and computes the similarity scores between two summaries. In this work, our experiments use the \textit{pre-trained universal-sentence-encoder-large} model \cite{cer-etal-2018-universal}. 

\textbf{BLEU} \cite{papineni2002bleu} is a textual similarity metric that calculates the precision of n-grams in a translated sentence compared to a reference sentence, with a weighted brevity penalty to punish short translations. We use the standard BLEU score which provides a cumulative score of uni-, bi-, tri-, and quat-grams. 

\textbf{ROUGE} \cite{lin2004rouge} is a popular automatic evaluation metric that is recall-oriented. It computes the count of several overlapping units such as n-grams, word pairs, and sequences. ROUGE has several different variants from which we consider the ROUGE-L. 

\textbf{METEOR} \cite{banerjee2005meteor} is a metric based on the general concept of unigram matching, and it combines precision, recall, and a custom score determining the degree to which words are ordered correctly in the translation. 

\subsection{ Evaluation Results and Analysis}
\begin{table*}
\centering
\caption{Evaluation results of ChatPT and three DL-based models}
\label{tab:eval-cgpt}
\begin{tabular}{c|cccc|cc}
\hline
\multirow{2}{*}{Code Sum Methods} & \multicolumn{4}{c}{reference-based} & \multicolumn{2}{|c}{reference-free} \\ \cline{2-7} 
                                 & BLEU  & ROUGE & METEOR & USE        & InRate          & MRR             \\ \hline
NCS                              & 21.55 & 35.98 & 15.08  & 0.5198     & 31.84\%          & 0.5614          \\
DOME                             & 22.20 & 36.67 & 16.47  & 0.5460     & 24.99\%          & 0.6296          \\
SIT                              & 22.54 & 37.87 & 16.11  & 0.5634     & 23.65\%          & 0.6404          \\ \hline
Human References                 & -     & -     & -      & -          & 15.05\%          & 0.8165 \\ \hline
gpt-3.5-turbo                    & 15.24 & 33.07 & 16.17  & 0.6164     & 5.33\%           & 0.8733          \\
text-davinci-003                 & 17.49 & 34.69 & 15.38  & 0.6234     & 3.65\%           & 0.8826          \\ \hline
\end{tabular}
\end{table*}

The semantic similarity based metric USE reveals a greater discrepancy among the NCS, DOME, and SIT models, highlighting the limitations of traditional n-gram overlap-based metrics like BLEU, ROUGE, and METEOR in capturing subtle semantic differences. 
This is evident from the initial three rows of Table~\ref{tab:eval-cgpt}, where the performance of NCS, DOME, and SIT models appears closely matched, with 
no more than 2\% difference
in BLEU, ROUGE, and METEOR scores. Such marginal disparities challenge the intuitive differentiation of model efficacy.
The traditional n-gram overlap-based metrics merely measure the literal proximity of predicted comments to reference comments, yet many words have close synonyms and certain words within a sentence carry more weight than others~\cite{haque2022semantic}. 
While comments generated by these three DL-based models may seem similar on a superficial literal level, the USE metric, which assesses semantic similarity through embedding vectors, provides a nuanced ability to distinguish the quality differences between the models' generated comments.

Our proposed reference-free metrics, IncRate and MRR scores, align with the four traditional reference-based metrics in evaluating three DL-based models. It is observed that SIT generated comments exhibit the lowest IncRate, DOME performs the second and NCS generated comments get the highest inconsistency rate with their corresponding code snippets. Simultaneously, SIT generated comments also achieve the highest MRR scores when they are used as code search queries among three DL-based baselines, while DOME ranks second and CSN performs the lowest MRR score. Furthermore, our proposed two reference-free metrics can reveal more substantial performance differences among NCS, DOME, and SIT compared to reference-based metrics. While n-gram-based metrics show less than a 2\% difference and USE metrics less than 5\%, IncRate and MRR metrics demonstrate a wider gap of 8\%. This highlights the effectiveness of IncRate and MRR in detecting subtle but crucial differences in comment quality across the DL-based code summarization models.

In comparing two ChatGPT models, gpt-3.5-turbo and text-davinci-003, with three DL-based baseline models, it's clear that ChatGPT falls behind in BLEU, ROUGE and METEOR scores but surpasses all three DL-baselines in the USE metric. This discrepancy indicates that relying solely on reference-based metrics could yield unreliable or contradictory assessments. Notably, as shown in Table~\ref{tab:tlc-stat}, two ChatGPT models exhibit a richer vocabulary in their comments, 12$\%$-23$\%$ larger than human written reference comments. The DL-based comments share the same vocabulary with human references. This suggests that ChatGPT-generated comments preserve more semantic diversity in word/token selection. Therefore, the results underscore the importance of semantic measurement over literal overlap for a more comprehensive evaluation of ChatGPT's capabilities.

Based on the semantic similarity metric USE, ChatGPT-generated comments show better likeness with human reference comments than three DL-based baselines. According to our reference-free metrics, both two ChatGPT models achieve lower IncRate and higher MMR scores than all three DL-based baselines, which aligns well with the conclusion obtained by the USE metric. 
\newtcolorbox{RQ1-box}{colframe = gray!50!black}
\begin{RQ1-box}
\textbf{Answer to RQ1} 
The reference-free metrics, IncRate and MRR, are effective in assessing code comment quality as they not only align with reference-based metrics within homogeneous models but also reveal more substantial performance differences across various models.
\end{RQ1-box}

The USE is a reference-based metric, although it can measure the semantic similarity between predicted comments and human reference comments. 
An inherent flaw rooted in the reference-based metric is that since human references are treated as the gold standard, we cannot fairly compare the quality of human reference comments with model-predicted comments.

In contrast, IncRate and MRR excel as reference-free automatic evaluation metrics, which are independent from reference standards. Their focus on the semantic relationship between a comment and its corresponding code enables a direct assessment of comment quality. The IncRate and MRR provide objective approaches to evaluate comment quality, regardless of whether the comment is produced by summarization models, generated by ChatGPT, or written by human developers. 

As shown in Table~\ref{tab:eval-cgpt}, 15.05\% of human reference comments in the TLC test set are detected as semantically inconsistent by the inconsistency detection classifier, suggesting that non-standard practices of co-evolving and maintaining comments with code often exist in software development and evolution. 
In comparison, the IncRate for gpt-3.5-turbo and text-davinci-003 generated comments are markedly lower, at 5.33\% and 3.65\%, respectively. 
Additionally, in the code search task where comments serve as queries, ChatGPT-generated comments also surpass human references.
Specifically, gpt-3.5-turbo and text-davinci-003 achieve MRR scores of 0.8733 and 0.8826, outperforming the human reference score of 0.8165. 
The quality of comments generated by the two ChatGPT models, text-davinci-003 and gpt-3.5-turbo, shows a minor difference. The text-davinci-003 produces slightly superior comments, while it is ten times more expensive than gpt-3.5-turbo.

\newtcolorbox{RQ2-box}{colframe = gray!50!black}
\begin{RQ2-box}
\textbf{Answer to RQ2} 
The ChatGPT-generated comments possess higher quality than human reference comments by preserving lower inconsistency and higher semantic relevance to the corresponding codes.
\end{RQ2-box}
\section{Distilling ChatGPT for Code Intelligence Tasks}
\label{sec:distillation}
In the realm of pre-trained source code models, the quality of training data is fundamental to the model's performance. The PL-NL paired data is crucial for bridging the semantic gap between programming and natural language, providing context and descriptive insight that facilitate code-intelligence tasks. Section~\ref{sec:evaluation} reveals that ChatGPT-generated comments exhibit higher semantic relevance and lower inconsistency with the corresponding code snippets than human reference comments. These findings indicate that incorporating ChatGPT-generated comments could significantly enhance the training data quality, thereby motivating the exploration of rebuilding pre-training dataset for code intelligence tasks. Therefore, we ask the following research questions:

\textbf{RQ3: How does the pre-training data rebuilt by ChatGPT impact the performance of the downstream code intelligence tasks?} 

\subsection{Experiment Design}
To assess the impact of ChatGPT-rebuilt data on code intelligence tasks, we conduct experiments as follows:
First, we reconstruct a popularly used dataset by substituting human-written comments with ChatGPT-generated ones. 
This step ensures improved semantic consistency between the NL comments and PL code snippets within the dataset.
Then, the rebuilt dataset is utilized to pre-train a widely used source code model. After that, fine-tune the pre-trained model in downstream tasks and measure their performances. 
We finally evaluate the impact of ChatGPT-rebuilt pre-training data by comparing the outcomes of the model trained on the rebuilt dataset with model trained on original human-referenced dataset, both quantitatively and qualitatively.

\begin{table}
\centering
\caption{Statistics of Cgpt-CSN dataset}
\label{tab:csn-stat}
\begin{tabular}{c|cc}
\hline
PLs        & W/ ChatGPT-NL      & W/o NL  \\ \hline
Ruby       & 53,269    & 110,551  \\
JavaScript & 138,577   & 1,717,933 \\
Go         & 346,333   & 379,103  \\
Python     & 457,429   & 657,030  \\
Java       & 496,651   & 1,070,271 \\
PHP        & 578,072   & 398,058  \\ \hline
Total      & 2,070,331 & 4,332,946 \\ \hline
\end{tabular}
\end{table}

\subsubsection{Dataset and Pre-trained Model}
We rebuild the popular dataset CodeSearchNet~\cite{husain2019codesearchnet} by replacing the human reference comments with ChatGPT generated ones, noted as cgpt-CSN. The prompt we used is shown in Figure~\ref{fig-prompt}.
As the empirical finding in Section~\ref{sec:evaluation}, the quality of comments generated by text-davinci-003 and gpt-3.5-turbo shows a minor difference. The text-dacinci-003 is slightly better than gpt-3.5-turbo while it is ten times more expensive than gpt-3.5-turbo.
Due to considerations of OpenAI API cost and computation cost, we utilize gpt-3.5-turbo for rebuilding the datast, and we did not incorporate the C/CSharp dataset added in the original CodeT5~\cite{wang2021codet5}. 
We utilized approximately 2.07 million paired PL-NL instances, encompassing six PL: Java, Python, PHP, Javascript, Go, and Ruby. The statistics of the rebuilt CodeSearchNet cgpt-CSN are displayed in Table~\ref{tab:csn-stat}. 

In this study, we choose CodeT5~\cite{wang2021codet5}, a widely-used pre-trained model for source code based on an encoder-decoder framework similar to T5~\cite{raffel2020exploring}, as our pre-training model for downstream code intelligence tasks.
We follow Feng et al.~\cite{feng2020codebert} to employ CodeSearchNet~\cite{husain2019codesearchnet} to pre-train CodeT5, which consists of six PLs with both unimodal and bimodal data. 
It is trained with four pre-training tasks, including Masked Span Prediction (MSP) task, Identifier Tagging (IT), Masked Identifier Prediction (MIP), and Bimodal Dual Generation (BDG).  

As the pre-training implementation is not available, we re-implement the CodeT5 pre-training process based on Huggingface’s T5~\cite{raffel2020exploring} PyTorch implementation, and the size of model is 220M, same as CodeT5-base. We set the maximum source and target sequence lengths to be 512. We use the mixed precision of FP16 to accelerate the pre-training. We set the batch size to 48 and employed the peak learning rate of 2e-5 with linear decay. Following the settings in CodeT5~\cite{wang2021codet5}, we pre-train the model with the denoising objective for 100 epochs and bimodal dual training for further 50 epochs. 
In the fine-tuning phase, for the hyperparameters in the CodeXGLUE~\cite{lu2021codexglue} tasks, such as learning rate, training steps, and batch size, we follow their default settings.

\subsubsection{Code Intelligence Tasks and Metrics}

For the code intelligence tasks, we cover 4 generation and 1 understanding tasks in the CodeXGLUE benchmark~\cite{lu2021codexglue} and employ the provided public datasets and the same data splits following it for all these tasks. We first consider two cross-modal, text-to-code and code-to-text generation tasks, two code-to-code generation tasks and one code understanding task. 

\textbf{Code summarization} aims to summarize a function-level code snippet into natural language descriptions. The dataset consists of six PLs including Ruby, JavaScript, Go, Python, Java, and PHP from CodeSearchNet~\cite{husain2019codesearchnet}. 
We employ one reference-based metric USE~\cite{haque2022semantic} and one our newly proposed reference-free metric MRR to evaluate code summarization.

\textbf{Code generation} is a NL-PL task that generating a code snippet based on NL descriptions. We employ the Concode dataset~\cite{iyer2018mapping} in Java where the input contains both NL texts and class contexts, and the output is a Java function. We evaluate it with BLEU-4, exact match (EM) accuracy, and CodeBLEU~\cite{ren2020codebleu} that considers syntactic and semantic matches based on the code structure in addition to the n-gram match.

\begin{figure}
\centerline{\includegraphics[scale=0.35]{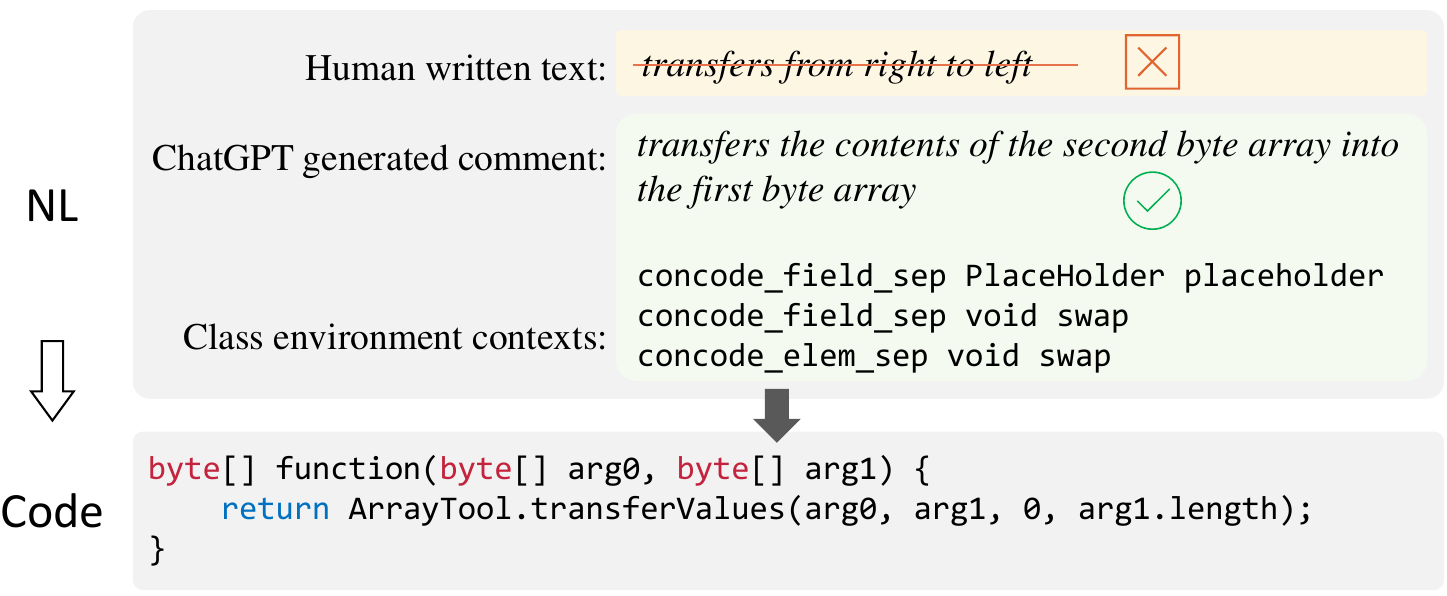}}
\caption{A Concode case rebuilt by ChatGPT.}
\label{fig-concode-rebuild}
\end{figure}

\textbf{Code translation} aims to migrate legacy software from one PL to another, where CodeXGLUE focus on translating functions from Java to CSharp and vice versa. Because we do not include C/CSharp in our updated version of pre-training dataset, we use the AVATAR~\cite{ahmad2021avatar}, a parallel corpus for Java-Python program translation, for code translation task. 

\textbf{Code refinement} is to detect which parts of code are buggy and fix them via generating a correct code sequence. We employ two Java datasets provided by Tufano et al.~\cite{tufano2019empirical} with various function lengths: small (fewer than 50 tokens) and medium (50-100 tokens). Due to the limit edit refining, there are large tokens overlap between source and target code. The EM measurement could better reflect the correctness of the refinement generation. We report EM and BLEU-4 scores for evaluation.  

\textbf{Clone detection} aims to measure the similarity between two code snippets and predict whether they have the same functionality. We experiment with the Java data provided by Wang et al.~\cite{wang2020detecting}. We employ F1 score and accuracy for evaluating these two tasks respectively. 

\subsection{Results and Analysis}

In this section, we valuate the re-pre-trained CodeT5 on downstream tasks. For better illustration, cgpt-CSN denotes the model was pre-trained with updated version of CodeSearchNet that we rebuilt using ChatGPT.

\begin{table*}
\caption{Evaluation results of Code Summarization}
\label{tab:eval-summarization}
\begin{tabular}{c|c|cccccccccccc}
\hline
\multirow{2}{*}{pre-train} & \multirow{2}{*}{fine-tune} & \multicolumn{2}{|c}{Javascript} & \multicolumn{2}{c}{PHP} & \multicolumn{2}{c}{Ruby} & \multicolumn{2}{c}{Java} & \multicolumn{2}{c}{Python} & \multicolumn{2}{c}{Go} \\
                           &                            & USE            & MRR           & USE        & MRR        & USE         & MRR        & USE         & MRR        & USE          & MRR         & USE        & MRR       \\ \hline
CSN                        & CSN                        & 0.5404         & 0.8102        & 0.6381     & 0.8102     & 0.5262      & 0.8189     & 0.6031      & 0.8083     & 0.5875       & 0.8294      & 0.6692     & 0.8580    \\
cgpt-CSN                   & CSN                        & 0.5420         & 0.8889        & 0.6429     & 0.8992     & 0.5375      & 0.8563     & 0.6111      & 0.9135     & 0.5911       & 0.9074      & 0.6763     & 0.9621    \\ 
CSN                        & cgpt-CSN-Sum                   & 0.7698         & 0.9623        & 0.8142     & 0.9202     & 0.7599      & 0.9477     & 0.8127      & 0.9342     & 0.6953       & 0.9529      & 0.8142     & 0.9120    \\
cgpt-CSN                   & cgpt-CSN-Sum                   & 0.7731         & 0.9665        & 0.8162     & 0.9617     & 0.7738      & 0.9617     & 0.8154      & 0.9553     & 0.6983       & 0.9552      & 0.8162     & 0.9244    \\ \hline
\multicolumn{2}{c|}{human references}            & -              & 0.7710        & -          & 0.8094     & -           & 0.7988     & -           & 0.8091     & -            & 0.8153      & -          & 0.8481    \\
\multicolumn{2}{c|}{gpt-3.5-turbo}           & -              & 0.9580        & -          & 0.9559     & -           & 0.9694     & -           & 0.9624     & -            & 0.9089      & -          & 0.9351    \\ \hline
\end{tabular}
\end{table*}

\textbf{Code Summarization} 
The initial two rows in Table~\ref{tab:eval-summarization} reveal that the model pre-trained with cgpt-CSN outperforms the one trained with vanilla CSN on code summarization task, as indicated by higher USE and MRR scores across all six PLs.
We further explore the effect of ChatGPT-generated comments data in fine-tuning phase, we rebuild the CSN code summarization datasets, denoted as cgpt-CSN-Sum. The results are shown in the middle two rows in Table~\ref{tab:eval-summarization}. It should be noted that the ground truth for reference-based metrics USE on rebuilt CSN summarization test set are ChatGPT generated comments. We observe that model fine-tuned with ChatGPT-rebuilt summarization data outperforms the model fine-tuned with human-referenced data.
These findings indicate that incorporating ChatGPT-generated data, both in pre-training and fine-tuning phases, advances code summarization performance.

The final two rows compare MRR scores between reference and ChatGPT-generated comments on the CSN summarization test set. ChatGPT-generated comments achieve markedly higher MRR scores across all six programming languages, aligning with findings from Section~\ref{sec:evaluation}. This consistency underscores the generalization capabilities across multiple programming languages.

\begin{table}
\centering
\caption{Evaluation results of NL-Code generation}
\label{tab:eval-concode}
\begin{tabular}{cc|ccc}
\hline
pre-train & fine-tune    & EM    & BLEU  & CodeBLEU \\ \hline
CSN       & Concode      & 22.00 & 38.52 & 39.45     \\
CSN       & cgpt-Concode & 29.60 & 48.31 & 49.25     \\
cgpt-CSN  & Concode      & 22.10 & 40.91 & 40.56     \\
cgpt-CSN  & cgpt-Concode & 30.00 & 50.20 & 50.49     \\ \hline
\end{tabular}
\end{table}

\textbf{Code Generation}
The vanilla input in Concode dataset contains both NL text and class environment context. 
To investigate the impact of ChatGPT-generated comments in fine-tuning stage, we also rebuild the Concode dataset by replacing the NL texts with ChatGPT-generated comments to form the updated inputs, as shown in Figure~\ref{fig-concode-rebuild}, and the updated version noted as cgpt-Concode. 
Table~\ref{tab:eval-concode} shows that the model pre-trained with cgpt-CSN and fine-tuned with cgpt-Concode outperforms other training-tuning settings in three metrics. Compared to the model pre-trained on CSN and fine-tuned on Concode, ChatGPT-generated comments data in both training and fine-tuning phases contributes to 8.0\%, 11.7\% and  11.0\% points improvement on EM, BLEU, and Code BLEU respectively. Particularly, we find that in the fine-tuning phase, the ChatGPT enhanced cgpt-Concode contributes greater performance gains than pre-training. 

We attribute this phenomenon to the characteristics of the NL-Code generation task itself. High-quality NL can better align the semantics with Code models, and the model learns better representations in the embedding space during the training process. 
In the NL-Code generation task, the quality of NL directly determines the semantic correlation between the input and the final target code. Therefore, in the fine-tuning stage, improving the NL quality of the fine-tuning data of the NL-Code generation task can significantly improve the quality of the generated code. 
These results demonstrate that high-quality comments data generated from ChatGPT could significantly advance the performance of the NL-code generation task.

\begin{table}
\centering
\caption{Evaluation results of Code Translation}
\label{tab:eval-translation}
\begin{tabular}{cc|ccc}
\hline
pre-train & fine-tune                    & CodeBLEU & BLEU  & EM   \\ \hline
CSN       & \multirow{2}{*}{Java2Python} & 48.28     & 51.29 & 2.57 \\
cgpt-CSN  &                              & 51.98     & 56.24 & 2.31 \\ \hline
CSN       & \multirow{2}{*}{Python2Java} & 55.52     & 56.82 & 1.21 \\
cgpt-CSN  &                              & 57.19     & 59.92 & 1.84 \\ \hline
\end{tabular}
\end{table}

\textbf{Code Translation} 
The experiment results of code translation task in the AVATAR testset are displayed in Table~\ref{tab:eval-translation}. We observe that the model pre-trained with cgpt-CSN data performs better than its counterpart pre-trained with human-commented CSN. Specifically, compared to the human-written comments, ChatGPT-generated comments in pre-training dataset achieve 3.7 and 1.67 CodeBLEU points higher performance in Java-to-Python and Python-to-Java translation, respectively. Somewhat surprisingly, code translation is an NL-unrelated task, and ChatGPT-generated NL in bimodal data can still improve the performance.

\begin{table}
\centering
\caption{Evaluation results of Code Refinement}
\label{tab:eval-refinement}
\begin{tabular}{cc|cc}
\hline
pre-train & fine-tune                       & EM    & BLEU  \\ \hline
CSN       & \multirow{2}{*}{Refine-small}  & 21.41 & 77.41 \\
cgpt-CSN  &                                & 21.24 & 77.36 \\ \hline
CSN       & \multirow{2}{*}{Refine-medium} & 13.90 & 89.39 \\
cgpt-CSN  &                                & 13.74 & 89.51 \\ \hline
\end{tabular}
\end{table}

\textbf{Code Refinement } Experiment results of EM and CodeBLEU scores for code refinement tasks are shown in Table~\ref{tab:eval-refinement}. The results indicate no significant performance differences between models trained with CSN and cgpt-CSN on both small and medium code refinement test sets. This implies that a deep comprehension of programming logic and syntax is essential for recognizing and correcting incorrect code patterns. The enhancement of NL comments in pre-training stage is less directly applicable to identifying and correcting code errors. 

\textbf{Clone Detection} Table~\ref{tab:eval-clone-detection} displays the results of F1 scores, Precision and Recall on clone detection task. No obvious F1-scores difference is found between the two models that were pre-trained with CSN/cgpt-CSN, respectively, as the performance difference is no more than 0.5 percentage point. The inputs of the clone detection task are two code snippets; the similarity between them is measured to predict whether they have the same functionality. We speculate that the similarity between code snippets is less about their natural language descriptions and more about their code content. Therefore, the ChatGPT-generated high-quality comments data does not directly transfer to better performance in recognizing code clones, which may depend more on code-specific features.

\begin{table}
\centering
\caption{Evaluation results of Clone Detection}
\label{tab:eval-clone-detection}
\begin{tabular}{cccc}
\hline
pre-train & F1     & P      & R      \\ \hline
CSN       & 0.9464 & 0.9455 & 0.9473 \\
cgpt-CSN  & 0.9432 & 0.9358 & 0.9508 \\ \hline
\end{tabular}
\end{table}

\newtcolorbox{RQ3-box}{colframe = gray!50!black}
\begin{RQ3-box}
\textbf{Answer to RQ3} 
For natural language related code intelligence tasks, such as code summarization and NL-code generation, high-quality comments pre-training data rebuilt by ChatGPT could enhance their performance.
\end{RQ3-box}

\section{Threats to validity}
\label{sec:threats}
\textbf{Prompt Variability in ChatGPT}. 
ChatGPT's outputs can vary significantly with slight changes in the prompt structure, wording, or context. A notable threat to the validity arises from the prompt we used to indicate ChatGPT for generating code comments. We employed a simple designed prompt across different programming languages to relieve this threat. While this method ensures easy replication and consistency across different code snippets, it does not account for the potential variability and nuanced responses that might emerge from more elaborately designed prompts. 

\textbf{Evolving Nature of ChatGPT}. 
A threat to the validity of our study stems from the versions of ChatGPT used in our research, specifically text-davinci-003 and gpt-3.5-turbo, which represent ChatGPT's capabilities at a certain point in time. 
ChatGPT is continuously updated and improved by its developers as a commercial product, leading to changes in its performance and output characteristics. While potentially beneficial, these updates can also introduce variability and unpredictability in the results. Therefore, our findings may have limited longevity.

\textbf{Limitations of Evaluation Metrics}. 
Another crucial threat to the validity of our study lies in the limitations of the evaluation metrics used for assessing code generation tasks. The metrics, like CodeBLEU, may not fully capture the quality and functionality of generated code in code intelligence tasks. However, these quantitative measures may overlook code correctness, potentially impacting the validity of our findings.

\textbf{Limitations in Generalizability}. 
The scope of our study's conclusions is constrained by limitations in computing resources and the expenses related to using the OpenAI API. Consequently, our research focused solely on a specific pre-trained code model, and we did not extend our analysis to include open-source Code Large Language Models such as Code Llama~\cite{roziere2023code}.

\section{Related works}
\textbf{ Quality Issue of Dataset}
Recent advancements in learning-based methods have significantly propelled the field of code intelligence tasks. As a data-driven paradigm, the quality of code-comment paired data critically determines these methods' effectiveness. 
Prior studies~\cite{linares2015developers, wen2019large, ibrahim2012relationship, malik2008understanding} have investigated the inconsistencies between code and comment from different perspectives. Fluri et al. found that 3-10\% of code comment changes lagged behind the corresponding code changes in seven Java open-source projects~\cite{fluri2007code}. 
Such obsolete comments may provide misleading information to developers, leading them to write vulnerable code~\cite{ibrahim2012relationship} and thus degrading the quality of the software. The noisy code comments data could degrade the performance of data-driven based learning models for code intelligence tasks~\cite{sun2022importance}.

The main focus of the research community is on developing customized models that can unleash the value of the available data in specific tasks. As mentioned in~\cite{zhao2021impact, liu2020simplifying}, improving the quality of the training data is still a research opportunity for machine learning, including DL-based source code models. Sun et al.~\cite{sun2022importance} proposed the first framework to improve the dataset quality for code search datasets. Their data cleaning framework, which consists of two subsequent filters: a rule-based syntactic filter and a model-based semantic filter, is considered to filter the noisy data. Xu et al.~\cite{xu2023data} investigated the data quality issue in the obsolete comment detection problem by proposing data cleaning and adversarial learning techniques. They found that the performance of DL models does improve with the cleaned training data. Unlike the above studies, our work tackles the data quality issue by rebuilding the dataset via LLMs, thereby replacing the noisy data with high-quality ones. 

\section{Conclusion}
We present a study comparing source code comments generated by ChatGPT, including  GPT-3.5-turbo and text-davinci-003, to human reference comments of the corresponding codes. To provide a fair and comprehensive comparison, we propose two auxiliary tasks inconsistency detection and code search as two reference-free metrics to evaluate their comments quality of. We found that ChatGPT-generated comments are superior to human written ones. This finding has two implications. First, the research community may reconsider using human-written references as the gold standard for training and evaluating source code models. Second, it could be promising to use ChatGPT to generate high quality data for advancing downstream code intelligence tasks.

This exploration motivates the reconstruction of the widely used CodeSearchNet dataset, wherein we substituted all natural language comments with those generated by ChatGPT. Following this, we employed this rebuilt dataset to re-pre-train the CodeT5 model. Subsequently, the pre-trained CodeT5 was fine-tuned on 4 generation and 1 understanding tasks. The outcomes of our experiments were revealing; the CodeT5 model, when trained with ChatGPT-generated data, notably outperformed its counterpart trained with human-generated comments across several NL-related tasks, including code summarization, code generation, and code translation. These results demonstrate the practicality of using ChatGPT to enhance the pre-training dataset for advancing code intelligence tasks.

\bibliographystyle{ieeetr}
\bibliography{IEEEexample}

\section{acknowledgement}
For a special day, 2016-12-19 to 2023-12-19.

\end{document}